\documentclass{Interspeech}
\usepackage{amsmath,graphicx, booktabs, multirow}
\usepackage{caption}
\usepackage{algorithm}
\usepackage{algpseudocode}
\usepackage{amssymb}
\usepackage{pifont}
\interspeechcameraready

\title{Clustering-based hard negative sampling for supervised contrastive speaker verification}

\author[affiliation={1,2}]{Piotr}{Masztalski}
\author[affiliation={1}]{Michał}{Romaniuk}
\author[affiliation={1}]{Jakub}{Żak}
\author[affiliation={1}]{Mateusz}{Matuszewski}
\author[affiliation={2}]{Konrad}{Kowalczyk}

\affiliation{}{Samsung R\&D Institute Poland} 
\ignorespaces \affiliation{}{AGH University of Krakow}{Poland}
\email{\{p.masztalski,m.romaniuk2,j.zak,m.matuszews2\}@samsung.com, konrad.kowalczyk@agh.edu.pl}
\keywords{speaker verification, supervised contrastive learning, hard negative sampling, clustering}

\usepackage{comment}

\begin{document}

\maketitle

\begin{abstract}
    
    In speaker verification, contrastive learning is gaining popularity as an alternative to the traditionally used classification-based approaches. Contrastive methods can benefit from an effective use of hard negative pairs, which are different-class samples particularly challenging for a verification model due to their similarity. In this paper, we propose CHNS - a clustering-based hard negative sampling method, dedicated for supervised contrastive speaker representation learning. Our approach clusters embeddings of similar speakers, and adjusts batch composition to obtain an optimal ratio of hard and easy negatives during contrastive loss calculation. Experimental evaluation shows that CHNS outperforms a baseline supervised contrastive approach with and without loss-based hard negative sampling, as well as a state-of-the-art classification-based approach to speaker verification by as much as 18 \% relative EER and minDCF on the VoxCeleb dataset using two lightweight model architectures.
\end{abstract}

\section{Introduction}
\label{sec:intro}

Speaker verification (SV) is a task that aims to verify the identity of a speaker based on voice characteristics. In recent years, researchers have been improving the performance of SV systems using various deep learning approaches with great success \cite{dnn-embeddings, dense-sv, xvector, ecapa, wavlm}. The standard method involves learning a fixed-size speaker embedding, which is used as a reference during pairwise comparisons in the evaluation process. The majority of current state-of-the-art systems are trained in a classification-based fashion, where during inference, the classification layer is discarded, and embeddings from the encoder are used for scoring. These methods are often trained with the Additive Angular Margin Softmax loss function (AAMSoftmax) introduced initially for face verification \cite{arcface}. 
While the classification-based approach is the most prevalent in the literature, there also exist multiple works that show the benefit of using contrastive learning or metric learning for SV, some of which report similar or better performance than the classification-based counterpart \cite{supcon_sv, supcon_sv_2, metric}. 
The contrastive approach to speaker verification has been explored primarily in self-supervised learning (SSL) \cite{simclr_sv, contrastive_sv, clustering_contrastive1, clustering_contrastive2, contrastive_margin}. Most SSL approaches are based on the SimCLR \cite{simclr} framework adapted from image representation learning, which maximizes agreement between differently augmented views of the same sample (positive pairs), while at the same time maximizes the distance between different samples (negative pairs). The training objective is achieved using the normalized temperature-scaled cross-entropy (NTXent) loss, which contrasts the similarity of a positive pair against the similarities between the anchor and negative examples in a batch. Due to the lack of class labels in SSL-based SV, positive pairs are typically constructed by augmenting the same audio utterance, which can result in the model learning channel characteristics (e.g. recording device, acoustic environment) instead of voice-dependent features \cite{simclr_sv}. Furthermore, a necessary assumption for SSL is that one always obtains segments of two different speakers when randomly sampling two audio utterances, which is not always true and can result in injecting false negatives into the loss calculation. 

Supervised contrastive learning \cite{supcon} is a natural extension of SimCLR to labeled datasets, and has also been used in the context of speaker verification, however, to a much lesser extent \cite{supcon_sv, supcon_sv_2, new_supcon}. In this approach, class labels are used to obtain diverse positive pairs that originate from different utterances of the same speaker, which forces the model to focus strictly on speaker-dependent audio features. Negative pairs are also selected based on speaker labels, mitigating the chance for any false negatives. Supervised contrastive learning utilizes the SupCon \cite{supcon} loss function, which is analogous to the NTXent loss with a simple modification to take class labels into account.

The contrastive learning approach seems to be intuitive for SV since it directly replicates the inference process during training, with pairwise embedding comparisons using cosine similarity. We also note an additional benefit of not requiring a classification layer, which can have a great impact on the parameter count of the model during training, when dealing with datasets containing a very large number of speakers. 

It has been shown in \cite{hcl, shcl, simcse, new_supcon} that contrastive learning can benefit from emphasizing hard negative samples. A hard negative is defined as a data point that belongs to a different class than an anchor sample, while at the same time producing a high similarity score. 
The standard approach to hard negative sampling for contrastive learning relies on incorporating a hardening function into the loss, so that negatives with higher similarity to the anchor sample have a larger influence on model updates than easy negatives. In supervised contrastive learning, this loss-based hard negative sampling method has been introduced in \cite{shcl} as H-SCL. 

In SV, hard negative sample pairs are utterances from different speakers with similar voice traits. A real-life example of that are speech samples from members of the same family \cite{family_voice}. This scenario is commonly found in a smart home setting, where a voice assistant should have a personalized approach based on the speaker identity, which makes it crucial for SV systems to be able to differentiate between similar speakers. The smart home setting also requires SV models to run efficiently on edge devices like smart vacuum cleaners, fridges or air conditioning units. Motivated by this fact, in our experiments, we consider lightweight model architectures suitable for real-time deployment on edge devices with limited compute capabilites.

Recent works focusing on contrastive and non-contrastive self-supervised speaker verification have shown that a clustering approach can be used to discover pseudo-labels in the training dataset \cite{clustering_contrastive1, clustering_contrastive2, clustering_dino1, clustering_dino2, domain_adaptation}. After a pre-training phase, the model is used to compute embeddings for training utterances, and cluster these embeddings into groups of similar samples. These groups are used to form positive pairs originating from different audio utterances, which results in improved SV performance compared to only augmenting the same segment.

While there is no need to use pseudo-labels in a supervised setting, in this paper we adapt the clustering method to find hard negative pairs and use them in the training process. Building on that idea, we introduce CHNS (\textbf{c}lustering-based \textbf{h}ard \textbf{n}egative \textbf{s}ampling), a hard negative sampling approach for supervised contrastive speaker representation learning. Contrary to H-SCL, we do not alter the loss function, and instead focus solely on the batch composition. We scan our training dataset for speakers with similar voice characteristics and cluster them into distinct groups. Our algorithm leverages these clusters to tune the ratio of hard negative samples inside each batch, and as a result improve performance of SV models.

\section{Proposed Method}
\label{sec:method}

Our proposed solution uses supervised contrastive learning to train a model that produces utterance-level speaker representations. We use a contrastive loss function during the training process, which calculates the relationship between speaker representations on a within-batch basis. This means that batch composition has a significant impact on the training process. In this section, we describe our batch sampling approach in detail.

\subsection{Clustering-based hard negative sampling}
\label{ssec:proposed-approach}

In contrast to a standard way of selecting hard negative samples on an utterance level, in CHNS, we discover hard negatives on the speaker (class) level. For each speaker in the training dataset, we sample 10 utterances, calculate an embedding for each utterance using a pre-trained SV model, and then calculate a centroid from these 10 embeddings. We use a baseline supervised contrastive model trained with random sampling to obtain the initial embeddings. This process is analogous to speaker enrollment in a real-life SV system and provides an average voice representation for each speaker. We will refer to these speaker embedding centroids as voiceprints.

Having obtained a voiceprint for each speaker in the training dataset, we run the K-Means clustering algorithm on them to create disjoint sets of similar speakers. We select K-Means specifically, because it uses squared Euclidean distance as a similarity measure. There is a linear relationship between squared Euclidean distance and cosine similarity when operating on normalized vectors \cite{cos_euc}, and during inference, the calculated embeddings are compared using cosine similarity. 

We assume that if there is a similarity between speaker voiceprints in each cluster, all of the pairs created from utterances that belong to different within-cluster speakers can be considered hard negatives. We argue that running the clustering algorithm on the voiceprint level is superior to clustering individual utterance embeddings, as it takes much less time to calculate and memory to store the results.

\begin{figure}[]
    \centering
    \includegraphics[width=7.5cm]{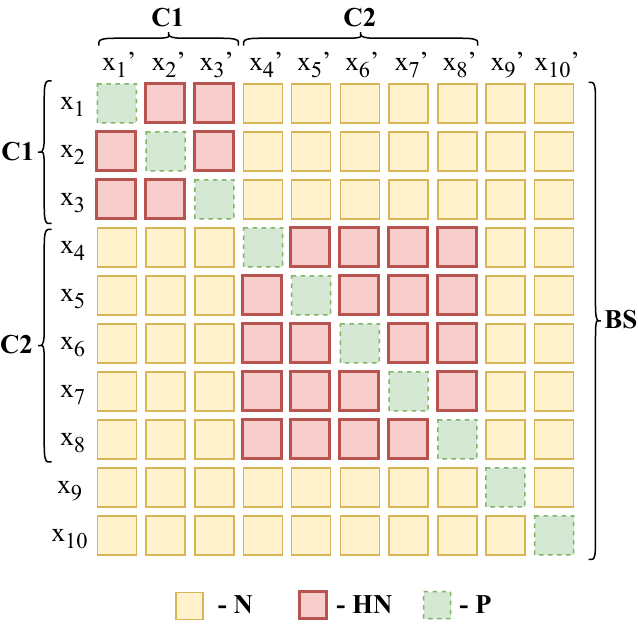}
    \caption{Proposed within-batch similarity matrix with \textit{hard ratio} set to 0.8. \textit{C1, C2} - cluster sizes, \textit{BS} - batch size, \textit{N} - negative, \textit{HN} - hard negative, \textit{P} - positive, \textit{(x\textsubscript{n}, x\textsubscript{n}')} - same speaker utterance pair.}
    \label{fig:cluster-batch}
    \vspace{-0.3cm}
\end{figure}

For the training process, we implement a custom batch sampler, which is designed to use the pre-calculated clusters for batch composition. First, we define a parameter called \textit{hard ratio}, which specifies what proportion of the batch is composed of speakers belonging to specific clusters. We start with an empty batch, and sample the first speaker cluster randomly. If a sampled cluster contains less than \textit{hard ratio}  $*$ \textit{batch size} speakers, another random cluster is sampled and its speakers are added to the batch. This process is repeated until we satisfy the \textit{hard ratio} condition. The remainder of the batch is sampled randomly from the rest of the training speakers. We sample two utterances per speaker to create a positive pair for each identity, and the remaining utterances in each batch form negative pairs. Using our algorithm, during the loss calculation we obtain a similarity matrix as depicted in Figure \ref{fig:cluster-batch}. The intra-cluster utterance pairs from different speakers form hard negatives, and the rest of the negative pairs are of random hardness. 


The number of clusters we choose for the K-Means algorithm will determine the average cluster size and is dependent on the number of speakers in the dataset. The larger the number of clusters, the smaller number of speakers belong to each cluster and the average distance of within-cluster speakers gets closer. This means that the number of clusters together with the value of \textit{hard ratio} plays a vital role in the final batch composition, and tuning these parameters to the training dataset is instrumental in obtaining optimal results.

\begin{table*}[t]
    \caption{Comparison between the number of clusters (groups of similar speakers) and different \textit{hard ratio} values on VoxCeleb1-H.}
    \vspace{-0.4cm}
    \label{tab:grid-search}
    \begin{center}
        \begin{tabular}{ccccccccc}
            \toprule
            \multirow{2}{*}{\textbf{\textit{Hard ratio}}}
                       & \multicolumn{2}{c}{\textbf{10 clusters}}         & \multicolumn{2}{c}{\textbf{20 clusters}}         & \multicolumn{2}{c}{\textbf{50 clusters}} & \multicolumn{2}{c}{\textbf{100 clusters}}         \\
            \cmidrule(lr){2-3}
            \cmidrule(lr){4-5}
            \cmidrule(lr){6-7}
            \cmidrule(lr){8-9}
             & \textbf{EER{[}\%{]}} & \textbf{minDCF} & \textbf{EER{[}\%{]}} & \textbf{minDCF} & \textbf{EER{[}\%{]}} & \textbf{minDCF} & \textbf{EER{[}\%{]}} & \textbf{minDCF} \\
            \midrule
            0.2        & 3.04                 & 0.1839          & 2.90                 & 0.1807          & 3.02                 & 0.1831    & 3.15 &  0.1903     \\
            0.5        & 3.01                 & 0.1866          & 2.92                 & 0.1844          & 2.86                 & 0.1753 &  2.91  & 0.1797    \\
            0.8        & 2.92                 & 0.1822          & 2.86                  & 0.1777          & 2.79        & 0.1721 & 2.88 & 0.1745 \\
           1        & 2.85                 & 0.1754          & 2.89                  & 0.1801          & \textbf{2.70}        & \textbf{0.1635} & 2.83 & 0.1743 \\
            \bottomrule
        \end{tabular}
    \end{center}
    \vspace{-0.7cm}
\end{table*}

\subsection{Loss function}
\label{ssec:alteratives}

The objective function in our contrastive experiments is given by:
\begin{equation}
\label{eq:loss}
    \mathcal{L} = \sum_{i=1}^{K} \frac{-1}{|P_i|} \sum_{x^+ \in P_i} \log \frac{e^{s(x_i, x^+) / \tau}}{e^{s(x_i, x^+) / \tau} + \sum\limits_{x^- \in N_i} H e^{s(x_i, x^-) / \tau}}
\end{equation} where \( x_i \) is the anchor sample, \(P_i\) and \(N_i\) are sets of positive and negative samples \( x^+ \) and \( x^- \) corresponding to the anchor sample, \(s(x, y)\) is the cosine similarity function, \(H\) is the hardening function, \(\tau\) is the temperature parameter and \(K\) is the batch size. Following H-SCL \cite{shcl}, we employ the exponential hardening function given by \(H = e^{\beta s(x,y)}\), where \(\beta\) is a hyperparameter. For \(\beta = 0\) the loss function is equivalent to the SupCon loss from \cite{supcon} without any weights to the negatives in the batch. When using one positive pair per speaker and \(\beta=0\), \(\mathcal{L}\) is also equivalent to the N-pair loss \cite{npair}.

The SupCon loss is the baseline comparison for both CHNS and H-SCL approaches to hard negative sampling. We also test the compatibility of our CHNS method with the additional weighing of negatives that is present in the H-SCL loss. The experimental results and findings are presented in Sec. \ref{sec:results}.

\section{Experimental setup and datasets}
\label{sec:exp}

We conduct a comprehensive experimental evaluation to find the best parameters for the proposed CHNS algorithm, and compare our solution to existing training methods: a supervised contrastive approach with random batch sampling (SupCon), a supervised contrastive approach with loss-based hard negative sampling (H-SCL), and an industry standard classification-based approach with the AAMSoftmax loss function. For a fair comparison, in all experiments, we use the exact same model, data and training parameters. We do not use additional data augmentation or score normalization.

\subsection{Models}
\label{ssec:model}

In the majority of performed experiments, we use the ECAPA-TDNN architecture \cite{ecapa} with a reduced number of convolution channels (256 channels which yields 2~M model parameters). We decrease the number of parameters to satisfy our aim of obtaining a model suitable for deployment on edge devices. For the classification-based training, we add one linear classification layer on top, which will be discarded during inference. For the contrastive methods, we decide not to use a non-linear projection head, which is contrary to SimCLR \cite{simclr}, and instead calculate the contrastive loss directly on the outputs of the encoder. Omitting the projection head seems intuitive in SV, since we do not fine-tune the model on a different downstream task as done in SimCLR. We have also found this approach to provide the best results. In order to test the transferability of our training method to other architectures, we additionally perform experiments using a 1.4~M parameter Thin ResNet-34 implementation from \cite{metric}.

\subsection{Data}
\label{ssec:data}

We use the VoxCeleb2 \cite{voxceleb2} development dataset to train the models. The dataset consists of over 1~M multilingual utterances from 5994 speakers. We extract a 100-speaker subset of the training dataset for validation. The training segments are cropped to 3 seconds at a 16 kHz sampling rate and converted to mel spectrograms. For testing purposes, we use VoxCeleb1-H - a challenging split from the VoxCeleb1 dataset, consisting only of speaker pairs of the same gender and nationality. In addition, we use the CNCeleb(E) \cite{cnceleb} evaluation dataset to test the cross-dataset performance of the models trained on VoxCeleb. We do not finetune the models on the CNCeleb training dataset.

As part of one experiment, we train the models on an internal dataset consisting of 1 s utterances of the "\textit{Bixby}" phrase from over 120~k diverse speakers. We evaluate the performance of these models on an in-house evaluation dataset to which we will refer as Bixby Eval.

For evaluation we use a standard set of SV metrics, namely Equal Error Rate (EER) and minimum Detection Cost Fuction (minDCF) with $c_{miss} = c_{fa} = 1$ and $p_{target} = 0.05$.

\subsection{Training setup}
\label{ssec:training}

For the classification-based training, we use the AAMSoftmax loss with margin and scale parameters set to 0.2 and 30 respectively. In case of contrastive experiments we use loss from \eqref{eq:loss} with the \(\tau\) parameter set as trainable starting with a value of 0.1. When employing H-SCL we set the \(\beta\) parameter to 0.1. We arrive at this value experimentally, as it balances training stability and model performance. For contrastive training without CHNS, we require that each batch contains only one positive pair per speaker, as we find this approach produces the best results. For all experiments, we use the Adam optimizer, and a cosine decay learning rate schedule with warm-up with maximum learning rate set to 0.01. Every training is run on a single NVIDIA A100 GPU for 200 epochs with a batch size of 1300. We share the code for reproducing our experiments online\footnote{https://github.com/masztalskipiotr/chns}.

\section{Experiments and results}
\label{sec:results}

\begin{table*}[t]
    \caption{Comparison between AAMSoftmax, SupCon, H-SCL and CHNS on VoxCeleb1-H, CNCeleb(E) and Bixby Eval.}
    \vspace{-0.4cm}
    \label{tab:main-results}
    \begin{center}
        \begin{tabular}{lcccccc}
            \toprule
            \multirow{2}{2cm}{\centering \textbf{Training strategy}}
                                     & \multicolumn{2}{c}{\textbf{VoxCeleb1-H}}   & \multicolumn{2}{c}{\textbf{CNCeleb(E)}}  & \multicolumn{2}{c}{\textbf{Bixby Eval}}     \\
            \cmidrule(lr){2-3}
            \cmidrule(lr){4-5}
            \cmidrule(lr){6-7}
                             & \textbf{EER{[}\%{]}} & \textbf{minDCF}& \textbf{EER{[}\%{]}} & \textbf{minDCF} & \textbf{EER{[}\%{]}} & \textbf{minDCF} \\
            \midrule
            AAMSoftmax       &  3.19                   & 0.1912   & 21.88                  & 0.6784          & 1.99                    & 0.1226          \\
            SupCon           &  3.17                   & 0.1919    & 21.72                  & 0.7076          & 1.82                    & 0.1152        \\
            H-SCL \cite{shcl}          &  3.02                  & 0.1817    &  20.07                 & 0.6903          & 1.68                    & 0.1123       \\
            SupCon + CHNS (ours) &  2.70          & 0.1635 & 16.61         & 0.5969 & 1.59           & 0.1066 \\
            H-SCL + CHNS (ours) &  \textbf{2.60}          & \textbf{0.1569} & \textbf{16.26}     & \textbf{0.5874} & \textbf{1.33}        & \textbf{0.0947} \\
            \bottomrule
        \end{tabular}
    \end{center}
    \vspace{-0.7cm}
\end{table*}

\setlength{\tabcolsep}{4pt}

In the first experiment we establish the best parameters for CHNS. The two variables in the algorithm are the number of clusters (which also correlates with the average cluster size) and \textit{hard ratio}. We test 4 cluster numbers of 10, 20, 50 and 100, and 4 \textit{hard ratio} values: 0.2, 0.5, 0.8 and 1. We compare results on the VoxCeleb1-H evaluation dataset to choose the best parameter combination. The results of these comparisons are presented in Table \ref{tab:grid-search}, which shows that the best performance is obtained when setting the number of clusters to 50 and \textit{hard ratio} to 1. In case of 50 clusters, the average cluster size is around 117 speakers, since there are 5894 speakers in our training dataset. This results in around 5.5 clusters in each batch when using a batch size of 1300 with \textit{hard ratio} = 1 (650 unique speakers per batch as there are 2 samples per each speaker). With a lower number of clusters, the number of speakers per cluster rises, and a bigger overall number of hard negatives are inserted into each batch. However, this comes at a cost of lower diversity within the batch, and the hard negatives are easier, on average, than in case of smaller clusters. The 50-cluster variant with \textit{hard ratio} of 1 setting seems to strike the best balance between batch "hardness" and speaker variability in case of the VoxCeleb2 dataset.

\begin{figure}[]
    \centering
    \includegraphics[width=8cm]{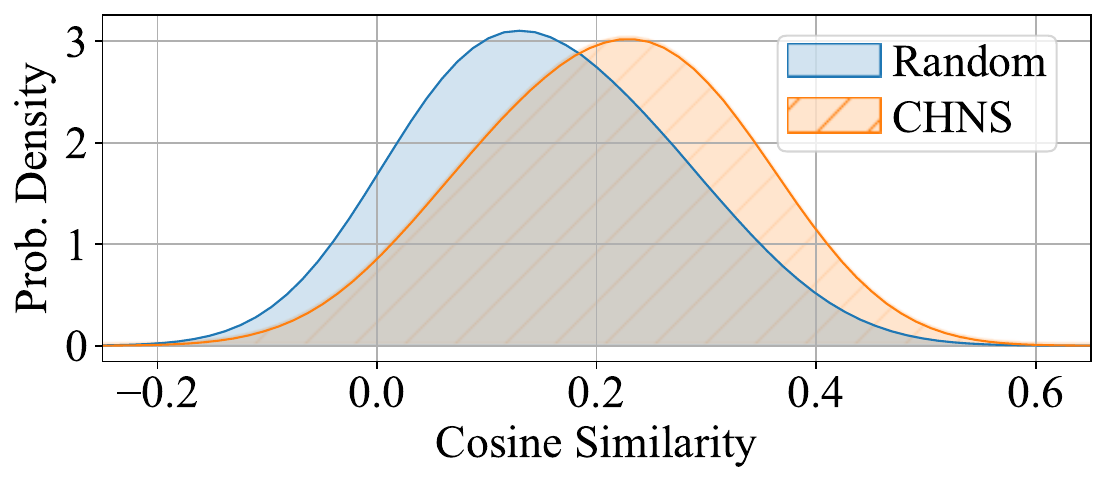}
    \caption{Average distribution of negative pair similarities in the training batch depending on the batch sampling method.}
    \label{fig:chns-vs-random}
    \vspace{-0.4cm}
\end{figure}

In the second experiment, we take the previously selected best model trained with CHNS (denoted as SupCon + CHNS) and compare it to the classification-based model (AAMSoftmax), a supervised contrastive model with random sampling (SupCon), and a supervised contrastive model with loss-based hard negative sampling (H-SCL) \cite{shcl}. To push the limits of hard negative sampling, as a last solution we also combine the H-SCL method with CHNS (denoted as H-SCL + CHNS). Table \ref{tab:main-results} presents the results of this experiment, which clearly indicate the usefulness of CHNS compared to the state-of-the-art on the three datasets. The comparison between default contrastive methods such as SupCon and H-SCL, and their corresponding CHNS versions (the proposed approach), shows that by changing only the batch composition method one can achieve significant performance improvements even though the training framework stays exactly the same. Additionally, when directly comparing the two hard negative sampling approaches, we observe that SupCon + CHNS significantly outperforms H-SCL, where the only difference between the two is the way of utilizing hard negative pairs. The advantage of CHNS is that it does actual \textit{sampling} of hard negatives instead of only assigning higher importance to some negative samples that are already in the batch, and therefore the average distribution of negatives within the batch skews more towards larger values compared to random sampling used in all competing methods (as shown in Figure \ref{fig:chns-vs-random}).  It is also worth noting that both contrastive methods which employ any form of hard negative sampling, usually outperform the classification-based AAMSoftmax model.


Looking at the results on CNCeleb(E) we can see a similar trend, as on VoxCeleb1-H, which indicates that the benefit of our method is transferable to other datasets. The larger absolute values of EER and minDCF are consistent with literature and are resulting from the lack of fine-tuning and the overall challenging nature of the CNCeleb dataset. 

When training and evaluating on our internal dataset, the HSCL + CHNS combination yields a relative improvement of 33 \% in EER and 23 \% in minDCF over the classification-based approach. We also note a meaningful advantage of using contrastive learning when training on datasets with a very large number of classes which is the lack of a classification layer. When training on our internal dataset using the AAMSoftmax loss, the classification layer alone contains over 23 M parameters which is more than 10 times the size of the applied ECAPA-TDNN encoder, and leads to much slower training times.

\begin{table}[]
    \caption{CHNS initialization step vs. results on VoxCeleb1-H.} 
    \vspace{-0.4cm}
    \label{tab:curriculum}
    \begin{center}
        \begin{tabular}{lcc}
        \toprule
        \textbf{Curriculum}     & \textbf{EER{[}\%{]}}   & \textbf{minDCF} \\
        \midrule
        From start              & \textbf{2.70} & \textbf{0.1635} \\
        After 5 \% of training  & 2.88          & 0.1746          \\
        After 50 \% of training & 2.94          & 0.1833         \\
        \bottomrule
        \end{tabular}
    \end{center}
    \vspace{-0.5cm}
\end{table}

For the next experiment, we implement a curriculum learning strategy that starts with a basic SupCon approach and gradually incorporates CHNS at different stages of training of the ECAPA-TDNN model. Specifically, we compare our default approach of using CHNS throughout the entire training process to introducing it after 10 epochs (5\% of the total training time) and after 100 epochs (50 \% of the total training time). For the curriculum learning approaches, we calculate the clusters for CHNS using the models trained for 10 or 100 epochs, thus avoiding the need for a pre-trained model to calculate voiceprint clusters before training. The results provided in Table \ref{tab:curriculum} indicate that starting CHNS training from the beginning yields the best performance. However, they also suggest that a model after just 5 \% of the total training time is sufficient to produce the embeddings required to calculate the voiceprints in our method. 

In the final experiment, we verify the model agnostic properties of CHNS, by comparing different training strategies on another lightweight model, i.e. Thin ResNet-34 \cite{metric}. We train the Thin ResNet-34 encoder on the VoxCeleb2 training dataset and test on VoxCeleb1-H. The results presented in Table \ref{tab:resnet-results} confirm a clear advantage of using the proposed CHNS approach with another popular architecture for speaker verification.

\begin{table}[]
    \caption{Comparison between training strategies with Thin ResNet-34 on VoxCeleb1-H.}
    \vspace{-0.4cm}
    \label{tab:resnet-results}
    \begin{center}
        \begin{tabular}{lcc}
        \toprule
        \textbf{Training strategy} & \textbf{EER{[}\%{]}}  & \textbf{minDCF} \\
        \midrule
        AAMSoftmax                 & 3.17          & 0.1882          \\
        SupCon                     & 3.42          & 0.2107          \\
        H-SCL \cite{shcl}                  & 3.25          & 0.2006          \\
        SupCon + CHNS (ours)           & 2.94 & 0.1845 \\
        H-SCL + CHNS (ours)           & \textbf{2.88} & \textbf{0.1784} \\
        \bottomrule
        \end{tabular}
    \end{center}
    \vspace{-0.9cm}
\end{table}

\section{Conclusion}
\label{sec:conclusion}

In this paper, we proposed CHNS - a clustering-based hard negative sampling approach for supervised contrastive speaker verification. Through a series of experiments using lightweight models suitable for use on edge devices, we demonstrated that our method outperforms standard supervised contrastive learning with and without loss-based hard negative sampling, as well as a state-of-the-art classification based approach to speaker verification by as much as 18 \% relative EER and minDCF, while being dataset and model agnostic. We open-sourced our training code.

\bibliographystyle{IEEEtran}
\bibliography{refs}

\end{document}